\documentclass[reprint,amsmath,amssymb,aps,pra,onecolumn]{revtex4-2}
\usepackage{graphicx}
\usepackage{amsmath}
\usepackage{amssymb}
\usepackage{amsthm}
\usepackage{braket}
\usepackage{hyperref}
\usepackage{quantikz}

\newtheorem{theorem}{Theorem}[section]

\begin{document}
\title{Quantum State Preparation for Complex-Valued Amplitudes}
\author{Fred Sun${}^{1,2}$}
\email[]{f4sun@uwaterloo.ca}
\affiliation{${}^{1}$softwareQ inc., Waterloo, Ontario, N2L 0C7, Canada}
\affiliation{${}^{2}$University of Waterloo, Waterloo, Ontario, N2L 3G1, Canada}
\date{\today}

\begin{abstract}
We present a state preparation procedure that takes binary registers containing the real and imaginary parts of a complex number $z=x+iy$ and prepares an amplitude approximating $z/\alpha$, where $\alpha$ is a normalization constant. The magnitude and argument are computed using the Coordinate Rotation Digital Computer (CORDIC) algorithm. We then encode the magnitude into an amplitude, and apply phase gates to introduce the argument. A reversible in-place multiplier allows the coordinate updates to use fixed precision while clearing their temporary workspace. The procedure uses $O(n^2)$ $T$ gates and $O(n)$ additional qubits. Compared with separately encoding the real and imaginary parts and combining them through LCU, the construction can allow a smaller normalization and reduce amplitude-amplification costs. The computed polar representation also remains available for further arithmetic before amplitude-encoding.
\end{abstract}

\maketitle

\section{Introduction}\label{sec:introduction}

Quantum computing has attracted significant interest as a framework for approaching problems in scientific computing that are computationally hard on classical computers, including Hamiltonian simulation, ground-state energy estimation, and many-body physics \cite{Martyn_2021, Lee_2021, Babbush_2018, Jordan_2012, Cao_2019}. These algorithms often require numerical values to be encoded into the amplitudes of a quantum state. However, values computed using quantum arithmetic are stored as binary digits in registers. In this work, we describe a state preparation procedure that takes the real and imaginary parts of a complex number stored in this form and encodes the number into an amplitude.\\

A complex number, $z=x+iy$, can be stored in two quantum registers containing the binary digits of its real and imaginary parts. This representation allows us to perform arithmetic on the stored values, but does not directly represent $z$ as an amplitude. We therefore consider how to convert these stored values into an amplitude while preserving the original inputs. The procedure must also work coherently when the inputs are in superposition, so that each input component is assigned its corresponding complex amplitude.\\

Given two registers $\ket{x}$ and $\ket{y}$ representing
$z=x+iy$, we would like to perform the transformation
\begin{align}
    \ket{x}\ket{y}\ket{0}\ket{0}_a
    \longmapsto
    \ket{x}\ket{y}
    \left(
        \frac{z'}{\alpha}\ket{0}+\ket{\perp}
    \right)\ket{0}_a,
    \label{eq:state_preparation_goal}
\end{align}
where $z'$ approximates $z$ and $\alpha$ is a normalization constant. Here, the third register is the output register, and $\ket{0}_a$ denotes all working ancilla registers. The state $\ket{\perp}$ is the remaining component perpendicular to the zero output state. It may depend on the input and is not necessarily normalized.\\

Our approach proceeds as follows. Writing $z=re^{i\phi}$,
we first compute approximations $r' \approx r$ and $\phi' \approx \phi$ using the Coordinate Rotation Digital Computer (CORDIC)
algorithm \cite{voldercordic,burge2026quantumcordicarcsine}. CORDIC computes the magnitude and argument through a sequence of bit-shifts and additions. This avoids expensive square-root and arctan operations when converting the input coordinates into magnitude and argument.\\

We then use the binary-to-amplitude encoding procedure \texttt{AMP}
to encode $r'/\alpha$
\cite{pocrnic2026efficientsimulationprebornoppenheimerdynamics}, and phase gates to introduce $e^{i\phi'}$. Together, these steps give the amplitude $z'/\alpha$, where $z'=r'e^{i\phi'}$. The stored magnitude and argument remain unchanged during encoding, allowing us to reverse the arithmetic and return all working ancillas to $\ket{0}$. \\

Our procedure may be useful within a larger algorithm, such as Hamiltonian simulation, where computing the complex inputs is expensive. Directly encoding the real and imaginary parts using \texttt{AMP} and combining them through LCU uses asymptotically fewer Toffoli gates, but our construction  can allow a smaller normalization than the LCU approach. This can reduce the number of amplitude-amplification steps and thus save resources when each step repeats expensive input computations. Furthermore, we explicitly compute the polar representation of $z$, allowing further arithmetic before amplitude-encoding. For example, this representation can be used in the fast-inversion procedure of Tong et al.~\cite{Tong_2021}: for nonzero $z=re^{i\phi}$, its reciprocal is $z^{-1}=r^{-1}e^{-i\phi}$, requiring a real reciprocal and negating the argument. \\

The remainder of the paper proceeds as follows. In Subsec.~\ref{sec:notation}, we introduce our notation and binary representation. In Sec.~\ref{sec:cordic}, we describe how to compute the magnitude and argument reversibly.
In Sec.~\ref{sec:state_preparation}, we use these values to prepare the desired amplitude and then uncompute the arithmetic registers.

\subsection{Notation}\label{sec:notation}

In this work, each input uses a fixed-point binary
representation with $n$ total bits: one sign bit,
$l$ integer bits, and $n-l-1$ fractional bits.
We write a number $a$ as
\[
    s_a\;a_{l-1}\dots a_0.a_{-1}\dots a_{-(n-l-1)}.
\]
We use two's complement, so its value is
\[
    a=-s_a2^l+\sum_{k=-(n-l-1)}^{l-1}a_k2^k.
\]
The register $\ket{a}$ contains these binary digits.
Arithmetic registers may include additional integer
and fractional bits to hold intermediate values
and provide the required precision.\\

We use $r'$ and $\phi'$ to approximate the magnitude and argument
of $z$, respectively.
We choose $\alpha$ to be a power of two large enough that $r<\alpha$ and $0\leq r'<\alpha$ for all allowed inputs, including the magnitude approximation error. We use the smallest such power of two. Angles are modulo $2\pi$, and we set the magnitude and argument to zero for zero input.

\section{Computing the Magnitude and Argument}\label{sec:cordic}

We use the \emph{Coordinate Rotation Digital Computer}
(CORDIC) algorithm to compute the magnitude and argument
of $z=x+iy$ using shifts and additions \cite{voldercordic}.
The resulting registers are used by the state preparation
procedure in Sec.~\ref{sec:state_preparation}.\\

Our goal is to compute approximations to
\[
    r=\sqrt{x^2+y^2},
    \qquad
    \phi=\operatorname{atan2}(y,x).
\]
CORDIC rotates the vector $(x,y)$ toward the positive
$x$-axis through a sequence of smaller rotations.
The accumulated angle gives the argument, while the
final horizontal coordinate gives the magnitude after
a known scaling correction.

\subsection{Preparing the Input}\label{subsec:quadrant}

We treat the real and imaginary parts of $z$ as the coordinates
of the vector $(x,y)$ and would like to compute its magnitude and argument.\\

To do so, we perform the CORDIC iterations on separate working
registers, leaving the original inputs unchanged.
We initialize the working registers by copying the
binary digits of $x$ and $y$ using CNOT gates:
\[
    \ket{x}\ket{y}\ket{0}\ket{0}
    \longmapsto
    \ket{x}\ket{y}\ket{x}\ket{y}.
\]
The new registers hold the working coordinates and include additional bits to accommodate intermediate values.
We keep the original inputs in order to uncompute the workspace after the amplitude-encoding procedure.\\

As a pre-processing step, we place the working vector in the right half-plane. This allows the same CORDIC iteration rule to be used for inputs in every quadrant. If $x<0$, we negate both working coordinates:
\[
    (x_0,y_0)=
    \begin{cases}
        (x,y), & x\geq 0,\\
        (-x,-y), & x<0.
    \end{cases}
\]
This preserves the magnitude but changes the argument by $\pi$. We account for this change by initializing the angle register to
\[
    \phi_0=
    \begin{cases}
        0, & x\geq 0,\\
        \pi, & x<0,\ y\geq 0,\\
        -\pi, & x<0,\ y<0.
    \end{cases}
\]
The CORDIC iterations then add the argument of the transformed vector to this initial correction, giving an approximation to the argument of the original input. Angle constants are stored to the chosen working precision. The original inputs are used to control the correction and to uncorrect it at the end of the procedure. For $x=y=0$, we set both outputs to zero.

\subsection{Cartesian-to-Polar CORDIC}

We perform $n$ iterations, indexed by $j=0,\ldots,n-1$.
At iteration $j$, we use the precomputed angle
\[
    \alpha_j=\arctan(2^{-j}).
\]
The vector is rotated clockwise when $y_j\geq 0$ and counterclockwise
when $y_j<0$. Writing the direction as
\[
    d_j=
    \begin{cases}
        -1, & y_j\geq 0,\\
        +1, & y_j<0,
    \end{cases}
\]
the ideal updates, before rounding, are as follows.
\begin{align}
    x_{j+1} &= x_j-d_j2^{-j}y_j,
    \label{eq:cordic_updates1}\\
    y_{j+1} &= y_j+d_j2^{-j}x_j,\notag\\
    \phi_{j+1} &= \phi_j-d_j\alpha_j.\notag
\end{align}
The angle update records the opposite of the rotation applied to the vector. Thus, it accumulates the original argument as the vector approaches the positive $x$-axis.\\

These operations don't include the normalization factor of a genuine rotation. Each iteration therefore increases the vector length by $\sqrt{1+2^{-2j}}$. After $n$ iterations, the correction factor is
\begin{align}
    K_n=
    \prod_{j=0}^{n-1}\frac{1}{\sqrt{1+2^{-2j}}}.
    \label{eq:km}
\end{align}
Consequently,
\[
    y_n\approx 0,
    \qquad
    K_nx_n\approx r,
    \qquad
    \phi_n\approx\phi.
\]
\subsection{Quantum CORDIC Implementation}
\label{subsec:quantum_cordic}

This procedure uses reversible ripple-carry adders \cite{cuccaro2004newquantumripplecarryaddition}.We denote arithmetic two's-complement conversion by \texttt{S2C}. 
Subtraction is implemented by performing an \texttt{S2C}, followed by an adder and an inverse of \texttt{S2C}.\\

In each iteration, before changing either coordinate, we copy the sign bit $s_j$ of $y_j$ into a fresh direction qubit. This qubit determines whether each update requires addition or subtraction, which can be done using an adder with one of the operands acted on by a controlled-\texttt{S2C}.
For a right-shift $x\gg j$ we approximate $2^{-j}x$ by rounding toward minus infinity.
We shift before applying the direction-dependent sign. The coordinate updates proceed as follows.\\

\begin{enumerate}
    \item We update the $x$ register:
    \[
        x_j\longmapsto x_{j+1}
        =x_j-d_j(y_j\gg j).
    \]
    The shifted operand is prepared from the unchanged $y$ register and is uncomputed after the addition or subtraction.\\

    \item We multiply the $y$ register by $1+2^{-2j}$ using Appendix~\ref{app:inplace_mult}:
    \[
        y_j\longmapsto y_j+(y_j\gg 2j).
    \]
    This gives the rounded product without adding fractional bits. Its temporary workspace is uncomputed.\\

    \item We add $d_j(x_{j+1}\gg j)$ to the modified $y$ register. To see why this works, consider the same steps without rounding:
    \begin{align}
        (1+2^{-2j})y_j+d_j2^{-j}x_{j+1}
        &=(1+2^{-2j})y_j
          +d_j2^{-j}(x_j-d_j2^{-j}y_j)\\
        &=y_j+d_j2^{-j}x_j,
    \end{align}
    This recovers the ideal $y$ update because $d_j^2=1$. The shifted operand is prepared from $x_{j+1}$ and uncomputed while that register is unchanged.
\end{enumerate}

This approximates both coordinate updates without storing a separate copy of $x_j$. The rounding error is bounded in Subsec.~\ref{subsubsec:polar_costs}.\\

We then update the angle register by adding $-d_j\alpha_j$, where $\alpha_j=\arctan(2^{-j})$.
We load $\alpha_j$ into a temporary register and conditionally negate it when $s_j=1$ using \texttt{S2C}. After adding the signed constant to the angle register, we reverse the conditional negation and unload the constant.\\

The temporary operand and carry registers are returned to $\ket{0}$ after use. The direction qubit is retained until the iteration is reversed during uncomputation after the whole procedure. At that point, $y_j$ has been restored, and its sign bit is used to return the direction qubit to $\ket{0}$. The sign bit is not uncomputed in each iteration and is left to uncompute the entire CORDIC procedure.\\

\begin{figure*}[t]
\centering
\resizebox{\textwidth}{!}{%
\begin{quantikz}[column sep=0.3cm, row sep=0.45cm]
\lstick{$\ket{0}_{s_j}$}
    & \targ{} & \qw\slice{1} & \ctrl{2} & \qw\slice{2} & \ctrl{2} & \qw & \qw\slice{3} & \qw
    & \octrl{2} & \qw\slice{4} & \octrl{2} & \qw\slice{5} & \ctrl{5} & \qw\slice{6} & \ctrl{5} & \qw\slice{7} & \rstick{$\ket{s_j}$}\\
\lstick{$\ket{x_j}$}
    & \qw & \qw & \qw & \gate[2]{\text{Add}_x} & \qw & \qw & \qw & \ctrl{1}
    & \qw & \qw & \qw & \ctrl{1} & \qw & \qw & \qw & \qw & \rstick{$\ket{x_{j+1}}$}\\
\lstick{$\ket{0}_{\mathrm{shift}}$}
    & \qw & \gate{\begin{gathered}\text{Copy \&}\\[-2pt]\text{Shift}_j\end{gathered}} & \gate{\text{S2C}} & \qw & \gate{\text{S2C}} & \gate{\begin{gathered}\text{Copy \&}\\[-2pt]\text{Shift}_j^\dagger\end{gathered}} & \qw & \gate{\begin{gathered}\text{Copy \&}\\[-2pt]\text{Shift}_j\end{gathered}}
    & \gate{\text{S2C}} & \gate[2]{\text{Add}_y} & \gate{\text{S2C}} & \gate{\begin{gathered}\text{Copy \&}\\[-2pt]\text{Shift}_j^\dagger\end{gathered}} & \qw & \qw & \qw & \qw & \rstick{$\ket{0}_{\mathrm{shift}}$}\\
\lstick{$\ket{y_j}$}
    & \ctrl{-3} & \ctrl{-1} & \qw & \qw & \qw & \ctrl{-1} & \gate{\begin{gathered}\text{Mult}\\[-2pt]1+2^{-2j}\end{gathered}} & \qw
    & \qw & \qw & \qw & \qw & \qw & \qw & \qw & \qw & \rstick{$\ket{y_{j+1}}$}\\
\lstick{$\ket{\phi_j}$}
    & \qw & \qw & \qw & \qw & \qw & \qw & \qw & \qw
    & \qw & \qw & \qw & \qw & \qw & \gate[2]{\text{Add}_\phi} & \qw & \qw & \rstick{$\ket{\phi_{j+1}}$}\\
\lstick{$\ket{0}_{\mathrm{ang}}$}
    & \gate{\begin{gathered}\text{Load}\\[-2pt]\alpha_j\end{gathered}} & \qw & \qw & \qw & \qw & \qw & \qw & \qw
    & \qw & \qw & \qw & \qw & \gate{\text{S2C}} & \qw & \gate{\text{S2C}} & \gate{\begin{gathered}\text{Unload}\\[-2pt]\alpha_j\end{gathered}} & \rstick{$\ket{0}_{\mathrm{ang}}$}
\end{quantikz}
}
\caption{One CORDIC iteration, with each slice marking the end of a stage.
(1) Copy the sign of $y_j$ into $s_j$, prepare $2^{-j}y_j$, and load $\alpha_j$.
(2) Use controlled \texttt{S2C} and addition to update $x_{j+1}$.
(3) Clear the shifted operand and multiply $y$ by $1+2^{-2j}$.
(4) Prepare $2^{-j}x_{j+1}$ and use controlled \texttt{S2C} and addition to update $y_{j+1}$.
(5) Clear the shifted operand.
(6) Add the signed angle $-d_j\alpha_j$ to $\phi$.
(7) Restore and unload the angle operand.
The \texttt{Copy \& Shift} gates copy the arithmetic right shift.}
\label{fig:cordic}
\end{figure*}
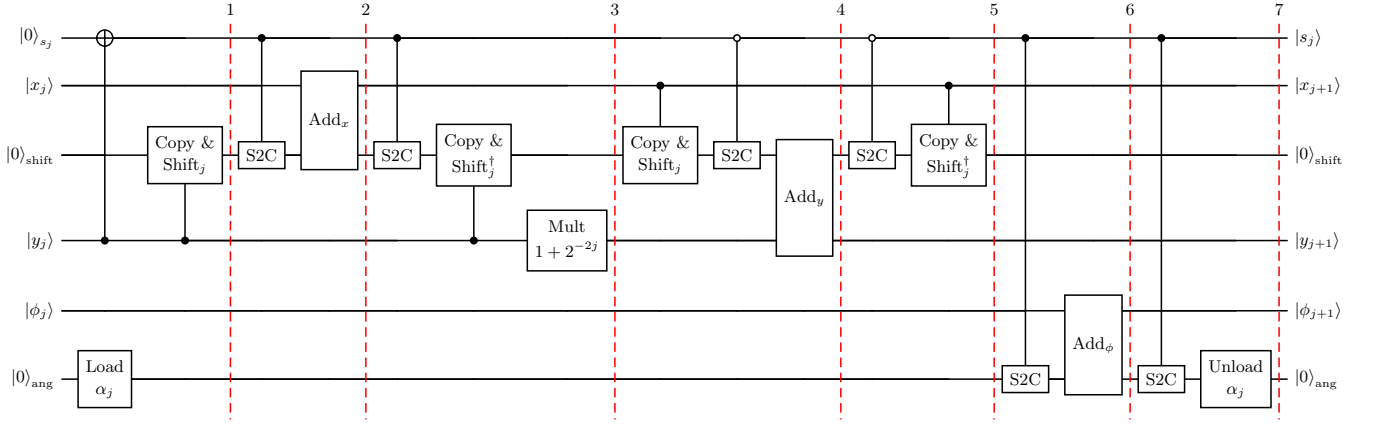

The circuit for one iteration is shown in Fig.~\ref{fig:cordic}.

\subsection{Recovering the Magnitude}

After the iterations, we multiply $x_n$ by the precomputed constant $K_n$ and store the result in a separate register:
\[
    \ket{x_n}\ket{0}
    \longmapsto
    \ket{x_n}\ket{r'},
    \qquad r'\approx K_nx_n.
\]
We compute $K_n$ classically to the required precision and load its binary approximation into an $O(n)$-qubit register. We then use a reversible multiplication circuit, chosen according to the available ancillas and desired circuit depth
\cite{litinski2024quantumschoolbookmultiplicationfewer,
sun2026polylogarithmicdepthquantummultiplier}.
For the resource bounds below, we use \cite{litinski2024quantumschoolbookmultiplicationfewer}.
We compute the full product, copy the required output bits into the magnitude register, and reverse the multiplication.
The unchanged $x_n$ and gain constant allow all intermediate product bits to be cleared.\\

The registers $\ket{r'}$ and $\ket{\phi'}$, with $\phi'=\phi_n$, are then used for magnitude and phase encoding in Sec.~\ref{sec:state_preparation}.

\subsection{Precision and Resource Costs}
\label{subsubsec:polar_costs}

We use fixed-point arithmetic throughout the iterations. Let $\delta$ denote the weight of the least significant bit in the working registers. 
To account for accumulated rounding error, we use $O(\log n)$ additional fractional bits, giving $\delta=O(2^{-(n-l)}/n)$.

\begin{theorem}\label{thm:precision}
If the input registers have least significant bit weight
$2^{-(n-l-1)}$ and the working registers use the
additional precision described above, then following
the construction in Subsec.~\ref{subsec:quantum_cordic}
gives a reconstructed complex value with error
$O(2^{-(n-l)})$.
\end{theorem}

\begin{proof}
For an arithmetic right shift by $k$ bits, we write
\[
    v\gg k=2^{-k}v+\epsilon_k,
    \qquad |\epsilon_k|<\delta.
\]
The error is zero when $k=0$.\\

First, consider the update to $x$. Since $d_j$ opposes the sign of $y_j$, we have
\begin{align}
    x_{j+1}=x_j+2^{-j}|y_j|+\epsilon_j,
    \qquad |\epsilon_j|<\delta.
\end{align}
Since the arithmetic right shift preserves the sign of the shifted operand, the implemented increment $-d_j(y_j\gg j)$ is nonnegative, and so $x_{j+1}\geq x_j$.\\

For the update of $y$, the implementation computes
\[
y_{j+1}=y_j+(y_j\gg 2j)+d_j(x_{j+1}\gg j).
\]
The in-place multiplication and the right shift each introduce an error of less than $\delta$. Substituting the rounded $x$ update and cancelling the $2^{-2j}y_j$ terms gives
\[
    y_{j+1}=y_j+d_j2^{-j}x_j+\eta_j,
    \qquad |\eta_j|<3\delta,
\]
where we account for the error coming from $x_{j+1}$. Since $d_j$ opposes the sign of $y_j$, it follows that
\begin{align}
    |y_{j+1}|
    \leq \bigl||y_j|-2^{-j}x_j\bigr|+3\delta.
\end{align}
We now bound the final $y$ value by induction. At the first iteration, the shifts are exact, so
\[
    |y_1|=\bigl||y_0|-x_0\bigr|
    \leq |y_0|+x_0=x_1.
\]
Suppose that
\[
    |y_j|\leq 2^{-(j-1)}x_j+3(j-1)\delta.
\]
Then
\[
    \bigl||y_j|-2^{-j}x_j\bigr|
    \leq 2^{-j}x_j+3(j-1)\delta.
\]
Using $x_{j+1}\geq x_j$, we obtain
\[
    |y_{j+1}|
    \leq 2^{-j}x_{j+1}+3j\delta.
\]
So, by induction,
\[
    |y_n|=O(2^{-n}x_n+n\delta).
\]

Thus, we introduce rounding errors of $O(\delta)$ when updating both $x$ and $y$ per iteration. Subsequent iterations multiply each error by at most a constant factor, giving a total error of $O(n\delta)$ \cite{kim2023erroranalysiscordicprocessor}.\\

We will now give an error bound on the approximate magnitude $r' \approx K_nx_n$ compared to the true $|z|$. The scaled vector length, $K_n\sqrt{x_n^2+y_n^2}$, differs from $|z|$ by $O(n\delta)$, and replacing it with $K_nx_n$ adds at most $K_n|y_n|$ error. Both inputs have magnitude at most $2^l$, and so the gain multiplication and the rounding error give $x_n=O(2^l+n\delta)$. Our bound on $y_n$ therefore gives an error of $O(2^{-n}2^l+n\delta)=O(2^{-(n-l)}+n\delta)$ when using the horizontal coordinate to approximate the magnitude. With our chosen precision, the total magnitude error is $O(2^{-(n-l)})$.\\

The remaining vertical component contributes at most $K_n|y_n|$ to the complex-value error, so the same bound applies. We store each angle constant to error $O(2^{-n}/n)$ and the gain constant to error $O(2^{-n})$. These errors and final magnitude rounding don't change the error bound, giving
\[
    |r'e^{i\phi'}-z|=O(2^{-(n-l)}),
\]
as claimed.
\end{proof}

Each iteration uses $O(n)$ Toffoli gates and $O(n)$ temporary qubits, which are cleared and reused. We also retain one direction qubit per iteration. The $n$ iterations and final multiplication use $O(n^2)$ Toffoli gates and $O(n)$ additional qubits. Reversing the arithmetic changes the gate count only by a constant factor.

\section{State Preparation Procedure}\label{sec:state_preparation}

We now use the computed registers $\ket{r'}$ and $\ket{\phi'}$
to prepare the amplitude in Eq.~\eqref{eq:state_preparation_goal}.
We encode the magnitude first and then introduce its phase.

\subsection{Encoding the Magnitude}\label{subsec:mag}

We encode the normalized magnitude $r'/\alpha$ into an amplitude using the \texttt{AMP} procedure by Pocrnic et al.~\cite{pocrnic2026efficientsimulationprebornoppenheimerdynamics}. Since $\alpha$ is a power of two, normalization is only a bitwise shift. We retain all stored bits and write
\begin{align}
        \frac{r'}{\alpha}=\sum_{j=1}^{m}b_j2^{-j},
    \qquad b_j\in\{0,1\},\qquad m=O(n).
\end{align}
\texttt{AMP} encodes this binary value while leaving the magnitude register unchanged:
\begin{align}
    \ket{r'}\ket{0}
    \longmapsto
    \ket{r'}
    \left(
        \frac{r'}{\alpha}\ket{0}
        +\ket{\perp}
    \right),
\end{align}
where $\ket{\perp}$ is the orthogonal component. The procedure exactly encodes the stored value, using $O(n)$ Toffoli gates and $O(n)$ additional qubits.

\subsection{Encoding the Phase}\label{subsec:arg}

We encode the argument by applying a phase gate to each qubit of $\ket{\phi'}$. For a bit with binary weight $2^j$, we apply
\[
    P(2^j)=\operatorname{diag}(1,e^{i2^j}).
\]
We use two integer bits in the argument register, so its sign bit has weight $-2^2$. We apply $P(-2^{2})$ on the sign bit of $\phi$ to account for its weight in the two's-complement representation. Together, these gates give
\begin{align}
    \ket{\phi'}\longmapsto e^{i\phi'}\ket{\phi'}.
\end{align}
Thus, each input component acquires its corresponding phase, and the desired output amplitude becomes
\begin{align}
    \frac{r'}{\alpha}e^{i\phi'}=\frac{z'}{\alpha}.
\end{align}
The stored argument remains unchanged.\\

We synthesize each of the $O(n)$ phase gates to error $O(2^{-n}/n)$. This gives a total synthesis error of $O(2^{-n})$ using $O(n^2)$ $T$ gates \cite{Kliuchnikov_2013}. Combining this with Theorem~\ref{thm:precision}, the total error in the desired amplitude is $O(2^{-(n-l)}/\alpha+2^{-n})$.

\subsection{Uncomputing the Arithmetic Registers}

The stored arithmetic values are preserved during amplitude and phase encoding. We first reverse the gain correction, then the CORDIC iterations in reverse order. Each inverse iteration restores $y_j$, whose sign bit clears its direction qubit.
Finally, we undo the quadrant correction and use the unchanged inputs to clear the initial coordinate copies.\\

With ideal phase gates, the final transformation is
\begin{align}
    \ket{x}\ket{y}\ket{0}\ket{0}_a
    \longmapsto
    \ket{x}\ket{y}
    \left(
        \frac{z'}{\alpha}\ket{0}
        +\ket{\perp}
    \right)\ket{0}_a,
\end{align}
where $\ket{\perp}$ is the component perpendicular
to $\ket{0}$ in the output register.

\subsection{Resource Costs}

With the precision and implementation in
Sec.~\ref{subsubsec:polar_costs}, computing the
magnitude and argument of $z$ requires $O(n^2)$
Toffoli gates and $O(n)$ ancillary qubits.
The \texttt{AMP} procedure adds $O(n)$ Toffoli gates
and output qubits. Uncomputing the arithmetic changes
its cost only by a constant factor.\\

The arithmetic and magnitude encoding therefore
require $O(n^2)$ Toffoli gates. Phase encoding
requires an additional $O(n^2)$ $T$ gates, and so the total procedure uses $O(n^2)$ $T$ gates. The total number
of additional qubits, including the output, is $O(n)$.

\section{Conclusion}

We have presented a state preparation procedure that takes
the real and imaginary parts of a complex number stored
in binary registers and encodes an approximation of that
number into an amplitude. We use CORDIC to compute the
magnitude and argument together through shifts and
additions, then encode the magnitude using \texttt{AMP}
and introduce the argument using phase gates.
The procedure acts coherently on superpositions of inputs.
The original inputs are preserved, and the arithmetic
is reversed after encoding to clear its temporary
registers.\\

This state preparation procedure may be useful within a larger algorithm
when a bound on the magnitude allows a smaller normalization
than separately encoding and combining the real and
imaginary parts. The resulting increase in success
probability can reduce the number of calls needed during
amplitude amplification. When these calls repeat expensive
input computations, the savings can outweigh the additional
CORDIC arithmetic. Comparing these costs in specific
applications is a natural next step.

\section{Acknowledgements}

The author gives thanks to Anton Borissov and Vlad Gheorghiu for the many insightful
discussions and feedback regarding this manuscript. This work was
supported in part by the National Research Council of Canada (NRC).

\section*{AI Disclosure}

This manuscript was initially drafted in March 2026 with limited assistance from OpenAI's ChatGPT for text editing. ChatGPT was used more extensively during revisions in October 2026 to assist with redrafting and editing the text and checking and formalizing mathematical arguments. The author takes responsibility for the final content and correctness of the manuscript.
\bibliography{complex_state_prep}

\appendix

\section{In-Place Multiplication by $1+2^{-2j}$}
\label{app:inplace_mult}

We implement
\[
    \ket{x}\ket{0}
    \longmapsto
    \ket{x'}\ket{0},
    \qquad x'\approx x(1+2^{-2j}),
\]
where $x'$ is the product rounded toward minus infinity
at the chosen fixed-point precision. We keep the same
number of fractional bits and allocate enough integer
bits to avoid overflow. All temporary registers return
to $\ket{0}$. We first consider $j>0$.

\subsection{Computing the Product}

The multiplication is addition with a shifted operand:
\[
   x(1+2^{-2j})=x+2^{-2j}x.
\]
We prepare a copy of $x$ and an arithmetic right-shifted
copy using CNOT gates, then add the shifted copy into
the first copy with a reversible ripple-carry adder
\cite{cuccaro2004newquantumripplecarryaddition} (we choose an input size such that there is no overflow).
The shift rounds down at the chosen
binary point.
Since the original input is unchanged, we reverse the
CNOT gates to clear the shifted copy. The adder clears
its carries. We then swap the input and product registers,
leaving
\[
    \ket{x'}\ket{x}.
\]
It remains to clear the old input.

\subsection{Recovering and Clearing the Input}

After the swap, we have the rounded product and the
old input. To clear the old input, we first reconstruct
it from the product alone. We then XOR the reconstructed
bits into the old input and reverse the reconstruction
circuit to clear its workspace.\\

The key point is that this particular rounded
multiplication still determines the input exactly.
Let $u$ and $v$ be the signed integers represented by
the input and product bits, with the same binary point,
and write $M=2^{2j}$. The shifted addition gives
\[
    v=u+\left\lfloor\frac{u}{M}\right\rfloor.
\]
To undo it, write $u=Mq+r$, where $0\leq r<M$.
The forward operation adds $q$, giving
\[
    v=(M+1)q+r.
\]
Dividing $v$ by $M+1$ therefore recovers the same
quotient $q$. Subtracting it recovers the input:
\[
    u=v-\left\lfloor\frac{v}{M+1}\right\rfloor.
\]
This identity also holds for negative inputs, using
rounding toward minus infinity.\\

We proceed with the recovery of $u$ using long division with blocks of $2j$ bits at a time following the long division process from left to right. Each remainder uses $2j+1$ bits, since division
by $M+1$ can produce remainders in the range $0\leq R\leq M$. For a nonnegative
dividend, suppose the current remainder is $R$ and the
next block has value $a$. Bringing down that block gives
$MR+a$ since the remainder has weight $M$ relative to the immediately less significant block. \\

Since
\[
    MR+a=(M+1)R+(a-R),
\]
we can take $R$ as the next quotient block and $a-R$ as the new remainder. If $a<R$ then the new remainder is negative. In this case, we take $R-1$ as the quotient block and add $M+1$ to the new remainder, giving
\[
    MR+a=(M+1)(R-1)+(a-R+M+1).
\]
Since $0\leq R\leq M$ and $0\leq a<M$, the quotient block always fits in $2j$ bits. If $a\geq R$, the
quotient block is $R\leq a<M$. If $a<R$, it is $R-1$, which lies between $0$ and $M-1$.
Thus, successive quotient blocks do not overlap.\\

The construction starts with $R=0$. For each input block $a$, we first compare $a$ with $R$ and store the result in bit $b$, where $b=1$ if $a<R$ and $b=0$ otherwise. We then compute the quotient block $R-b$ and the new remainder
$a-R+b(M+1)$, which can be computed using a controlled-add, in fresh registers. Thus, $b$ controls whether we subtract one from the quotient block
and add $M+1$ to the remainder. \\

We use the new remainder to process the next input block, while keeping the
previous remainder, the comparison bit $b$, and all working ancillas. This allows us to reverse the entire division after clearing the old input.\\

For $v<0$, we first use $|v|$ as the dividend. If the
remainder is zero, negating the quotient gives the required result. Otherwise, we also subtract one to round toward minus infinity.\\

We subtract the quotient from a copy of $v$ to obtain $\ket{u}$ in the workspace. We then apply CNOT gates
from the recovered bits to the old-input register, clearing it to $\ket{0}$. We reverse the subtraction, leaving the copy of $v$ in the workspace.
We clear this copy using CNOT gates from the unchanged product, then reverse the division to clear the quotient and all intermediate workspace.
This reversal works because recovery used only the product, which is unchanged; it never used the old-input register that we have just cleared.

\subsection{Resource Costs}

For $2j<n$, our division procedure uses $\lceil n/(2j)\rceil$ blocks, each requiring $O(j)$ Toffoli gates and ancillas. For $2j\geq n$, the quotient is $0$ for $v\geq0$ and $-1$ otherwise. Including the addition and uncomputation, we use $O(n)$ Toffoli gates and $O(n)$ clean ancillas for this procedure. All temporary registers return to $\ket{0}$.\\

For $j=0$, multiplication is exactly doubling. With one additional integer bit, we implement it
using only CNOT and SWAP gates and clear the old input using the shifted output.

\end{document}